\documentclass[10pt,prl,twocolumn,aps,superscriptaddress, longbibliography]{revtex4-1}
\usepackage{graphicx,amsmath,xcolor}
\usepackage[T1]{fontenc}
\usepackage{hyperref}
\usepackage{enumitem}
\usepackage{multirow}
\usepackage{quantikz}
\usepackage{textpos}
\usepackage{tikz-3dplot}
\usepackage[normalem]{ulem}
\makeatletter
\renewcommand\@makefnmark{\hbox{\textsuperscript{\normalfont\@thefnmark}}}
\makeatother

\begin{document}

    \title{Singlet-only Always-on Gapless Exchange Qubits with Baseband Control}
    \author{Nathan L. Foulk}
    \affiliation{Condensed Matter Theory Center and Joint Quantum Institute, Department of Physics, University of Maryland, College Park, Maryland 20742-4111 USA}
    \author{Silas Hoffman}
    \affiliation{Condensed Matter Theory Center and Joint Quantum Institute, Department of Physics, University of Maryland, College Park, Maryland 20742-4111 USA}
    \affiliation{Laboratory for Physical Sciences, 8050 Greenmead Drive, College Park, Maryland 20740, USA}
    \author{Katharina Laubscher}
    \affiliation{Condensed Matter Theory Center and Joint Quantum Institute, Department of Physics, University of Maryland, College Park, Maryland 20742-4111 USA}
    \author{Sankar Das Sarma}
    \affiliation{Condensed Matter Theory Center and Joint Quantum Institute, Department of Physics, University of Maryland, College Park, Maryland 20742-4111 USA}
\begin{abstract}
 We propose a singlet-only always-on gapless exchange (SAGE) spin qubit that encodes a single qubit in the spins of four electrons while allowing universal baseband control. While conventional exchange-only qubits suffer from magnetic-field-gradient-induced leakage and coherent errors due to local nuclear environments and variations in the $g$-factor, the SAGE qubit subspace is protected from coherent errors due to local magnetic field gradients and leakage out of the computational subspace is energetically suppressed due to the exchange interactions between electrons being always-on. Consequently, we find that when magnetic gradient noise dominates over charge noise, coherence times and single-qubit gate infidelities of the SAGE qubit improve by an order of magnitude compared to conventional exchange-only qubits. Moreover, using realistic parameters, two-qubit gates can be performed with a single interqubit exchange pulse with times comparable in duration to conventional exchange-only qubits but with a significantly simplified pulse sequence. 
\end{abstract}
\maketitle 

\paragraph{Introduction.}

Electrons localized in gate-defined semiconducting quantum dots are a promising candidate for building a large-scale fault-tolerant quantum computer. The most studied implementation is the so-called Loss-DiVincenzo (LD) qubit in which quantum information is encoded in the spin of a single electron~\cite{loss_quantum_1998, burkard_coupled_1999, burkard_semiconductor_2023,neyens_probing_2024,george_12-spin-qubit_2024,steinacker_300_2024,elsayed_low_2022}. Control of LD qubits typically necessitates ac electric fields (causing heating effects) and on-chip micromagnets. These experimental requisites have inhibited scale-up of LD qubits~\cite{philips_universal_2022,xue_computing_2021,noiri_fast_2021,wang_operating_2024,takeda_optimized_2018}.

Alternatively, exchange-only (EO) qubits use three electrons localized to three quantum dots to form a single encoded qubit. EO qubits are so-named because there exist electrically controllable exchange interactions between the spins, $H_\textrm{exch} = \frac{1}{4}\sum_{i<j} J_{ij} \boldsymbol{s}_i \cdot \boldsymbol{s}_j$, where $J_{ij}$ is the magnitude of the exchange interaction between spin $i$ and $j$ and $\boldsymbol{s}_j=(s_j^x,s_j^y,s_j^z)$ is a vector of the generators of rotation of the $j$th spin about $x$, $y$, and $z$ axis, respectively. These exchange interactions can be pulsed to access any point on the encoded qubit Bloch sphere. This all-electrical dc control and the absence of a micromagnet have drawn substantial experimental interest~\cite{acuna_coherent_2024,heinz_fast_2024,weinstein_universal_2023,ha_flexible_2022,andrews_quantifying_2019,kerckhoff_magnetic_2021,russ_quadrupolar_2018,bosco_exchange-only_2024}. Although EO qubits are immune to global magnetic fields, coherence times and gate fidelities are limited by differences in local magnetic fields induced by nuclear magnetic environments or differences in the $g$-factor between dots. Specifically, these effective magnetic field gradients drive both coherent errors within the qubit subspace and cause leakage to noncomputational states.

There exists a singlet-only~\cite{sala_exchange-only_2017, sala_highly_2020} encoding of four electrons across four quantum dots which is immune to magnetic-field-gradient-induced coherent errors within the qubit subspace~\cite{bacon_universal_2000,bacon_robustness_1999,bacon_encoded_2001,bacon_decoherence_2003,kempe_encoded_2001,kempe_theory_2001} while retaining the control advantages of conventional EO qubits. Four electrons coupled through the exchange interaction with Hamiltonian $H_\textrm{exch}$ can be used to define an encoded qubit whose states lie in the $S = S^z = 0$ subspace, where $S$ is the total spin quantum number and $S^z$ is the quantum number associated with the eigenvalue of the spin operator along the $z$-axis. The qubit states in the singlet-only encoding are
\begin{gather}
 \ket{0} = \ket{S_{12}S_{34}}, \\
 \ket{1} = \frac{1}{\sqrt{3}}\Bigl(\ket{T^0_{12}T^0_{34}} - \ket{T^+_{12}T^-_{34}}- \ket{T^-_{12}T^+_{34}}\Bigl),
\end{gather}
where $\ket{S_{ij}}$ refers to a singlet spin state and $\ket{T_{ij}^0}$, $\ket{T_{ij}^+}$, and $\ket{T_{ij}^-}$ refer to triplet spin $0$, $1$, and $-1$ states, respectively, on dots $i$ and $j$.
The four spins are coupled into pairs, and the qubit state is determined from the spin parity of these pairs. Without loss of generality, the spins of the first (second) encoding pair are labeled as 1 and 2 (3 and 4).

 Unfortunately, these singlet-only qubits are unprotected from leakage into noncomputational states driven by magnetic field gradients. Previous proposals for four-electron exchange qubits either did not address this problem~\cite{bacon_universal_2000,bacon_robustness_1999,bacon_encoded_2001,bacon_decoherence_2003,kempe_encoded_2001,kempe_theory_2001,hsieh_explicit_2003}, required all-to-all connectivity that would be difficult to manufacture~\cite{bacon_coherence-preserving_2001,weinstein_quantum-dot_2005,weinstein_scalable_2007,antonio_two-qubit_2013,jiang_preparation_2009}, or relied on ac driving of a gapped qubit~\cite{russ_quadrupolar_2018,sala_exchange-only_2017}, which reintroduces the ac-induced heating that prompted the switch to exchange qubits.

\paragraph{Qubit definition and coherence times.} We propose a singlet-only always-on gapless exchange-only (SAGE) spin qubit that provides protection from magnetic field gradients with scalable baseband control and suppressed leakage to noncomputational states. The SAGE qubit is implemented in a T-shape geometry as shown in Fig.~\ref{fig:1Q_layout_blochsphere}(a).
The resultant encoded qubit Hamiltonian is [see also the Supplemental Material (SM)~\cite{supplemental}]
\begin{equation}
 H_\textrm{q} =  \frac{J_{14}}{4} (\sqrt{3}\sigma^x + \sigma^z)
 - \frac{J_{13}}{4}(\sqrt{3}\sigma^x - \sigma^z) - \frac{J_{12}}{2} \sigma^z.
    \label{eq:qubit_ham}
\end{equation}
Any rotation in the $x$-$z$ plane can be performed by decreasing the appropriate exchange couplings [see Fig.~\ref{fig:1Q_layout_blochsphere}(b)]. Furthermore, when $J_{12} = J_{13} = J_{14}=J$, the qubit states are degenerate [see Eq.~(\ref{eq:qubit_ham})]. This always-on, gapless operation improves on previous four-spin exchange qubits by energetically suppressing leakage while preserving dc control in the non-rotating frame. The spectral gap between qubit states and leakage states is maximized in the T-geometry, see SM~\cite{supplemental}.

SAGE qubit measurement is straightforward using Pauli spin blockade (PSB)~\cite{kouwenhoven_quantum_1998, petta_coherent_2005, sala_exchange-only_2017, burkard_semiconductor_2023} so that electrons 1 and 2 are pulsed diabatically toward the same dot. A charge measurement then determines the qubit state~\cite{petta_coherent_2005} since the two electrons will only occupy the same dot if they form a spin singlet. 
This charge measurement is reliable due to the singlet-triplet relaxation time being much longer than the charge relaxation time~\cite{fujisawa_allowed_2002}. However, similar to PSB measurements in conventional EO devices, the measurement is sufficient only when the qubit state is unleaked. SAGE qubit initialization can be performed either by spin relaxation or by tunneling cold electrons into the singlet state from a neighboring reservoir~\cite{blumoff_fast_2022}.

\begin{figure}
    \scalebox{1.2}{
    \begin{minipage}{0.35\linewidth} 
    \hspace{-3mm}
    \begin{tikzpicture}
        \node at (0.5,2) {\fontsize{8pt}{12pt}\selectfont(a)};
        \node at (0.5, 4.75) {\textcolor{white}{(a)}};
        \node at (3 ,2.5) {\textcolor{white}{(a)}};
        \draw[red, line width=0.3mm] (1,3) -- (2,3) node[below left, xshift=-1.1mm]{$J_{13}$};
        \draw[blue, line width=0.3mm] (2,3) -- (3,3) node[below left, xshift=-1.1mm]{$J_{12}$};
        \draw[green!50!black, line width=0.3mm] (2,3) -- (2,4) node[below right, yshift=-2.5mm]{$J_{14}$};
        \node[circle, draw, fill=black, text centered] at (2, 3) {\textcolor{white}{1}};
        \node[circle, draw, fill=black, text centered] at (1, 3) {\textcolor{white}{3}};
        \node[circle, draw, fill=black, text centered] at (2, 4) {\textcolor{white}{4}};
        \node[circle, draw, fill=black, text centered] at (3, 3) {\textcolor{white}{2}};
    \end{tikzpicture}
    \end{minipage}
 }
    \hspace{2mm}
    \begin{minipage}{0.5\linewidth} 
    \tdplotsetmaincoords{70}{110}
    \begin{tikzpicture}[tdplot_main_coords]
        \node at (2.5,-1.2,-1) {(b)};
    
        \draw[->, opacity=0.8] (0,0,0) -- (2.0,0,0)  node[below right]{$y$};;
        \draw[->, opacity=0.8] (0,0,0) -- (0,-2,0) node[below left]{$x$};
        \draw[->, opacity=0.8] (0,0,0) -- (0,0,2.0) node[anchor=south]{$z$};
        \draw[->,opacity=0.6] (0,0,0) -- (-2.0,0,0);
        \draw[->,opacity=0.6] (0,0,0) -- (0,2,0);
        \draw[->,opacity=0.6] (0,0,0) -- (0,0,-2.0);
        \draw[dashed,opacity=0.4] (0,-1.73,-1) -- (0,-1.73,0);
        
        \draw[dashed,opacity=0.4] (0,1.73,-1) -- (0,1.73,0);
        \filldraw[black, opacity=0.5] (0,1.73,0) circle (0.5pt);    
        \filldraw[black, opacity=0.5] (0,-1.73,0) circle (0.5pt);    
        \shade[ball color=blue!10, opacity=0.3] (0,0,0) circle (2cm);

        \draw[line width=0.4mm,blue,->] (0,0,0) -- (0,0,2) node[below right]{$\vec{J_{12}}$};
        \draw[line width=0.4mm,green!50!black,->] (0,0,0) -- (0,1.73,-1) node[ below right]{$\vec{J_{14}}$};
        \draw[line width=0.4mm,red,->] (0,0,0) -- (0,-1.73,-1) node[below left]{$\vec{J_{13}}$};
        \foreach \p in {0} {
            \draw[dashed] plot[samples=100, domain=0:360] ({2*cos(\p)*cos(\x)},{2*cos(\p)*sin(\x)},{2*sin(\p)});
 }
        \foreach \p in {90} {
            \draw[ opacity=0.5] plot[samples=100, domain=0:360] ({2*cos(\x)*cos(\p},{2*cos(\x)*sin(\p)},{2*sin(\x});
 }
    
    \end{tikzpicture}
    \end{minipage}
    \caption{\textit{Qubit geometry and operation.} (a) Schematic of the SAGE qubit layout in real space. The encoding pairs of spins are $\{1,2\}$ and $\{3,4\}$. (b) Bloch sphere representation of the SAGE qubit and each exchange interaction's rotation axis of control.}
\label{fig:1Q_layout_blochsphere}
\end{figure}

\begin{figure}
    \centering
    \includegraphics[width=0.9\linewidth]{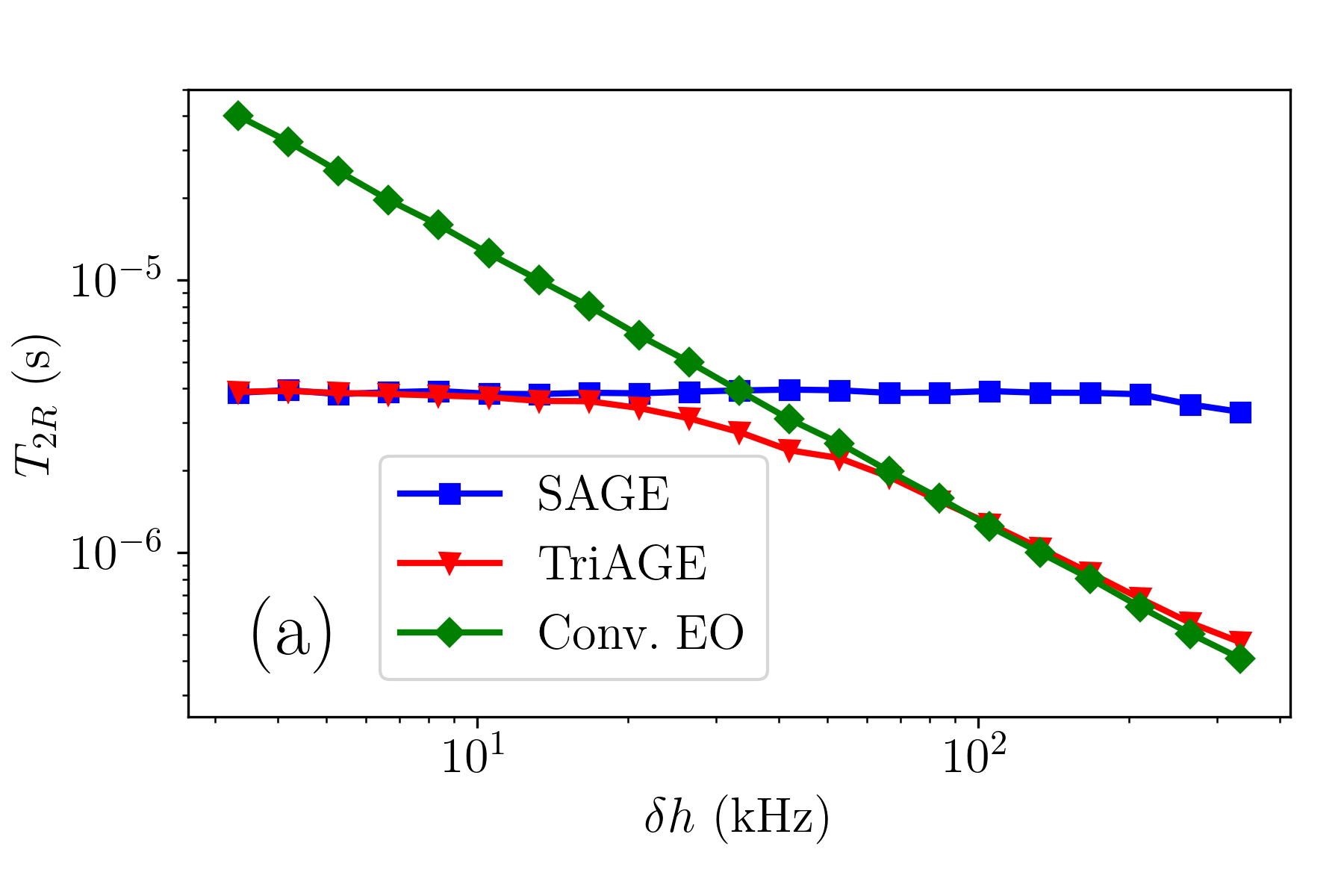}
    \includegraphics[width=0.9\linewidth]{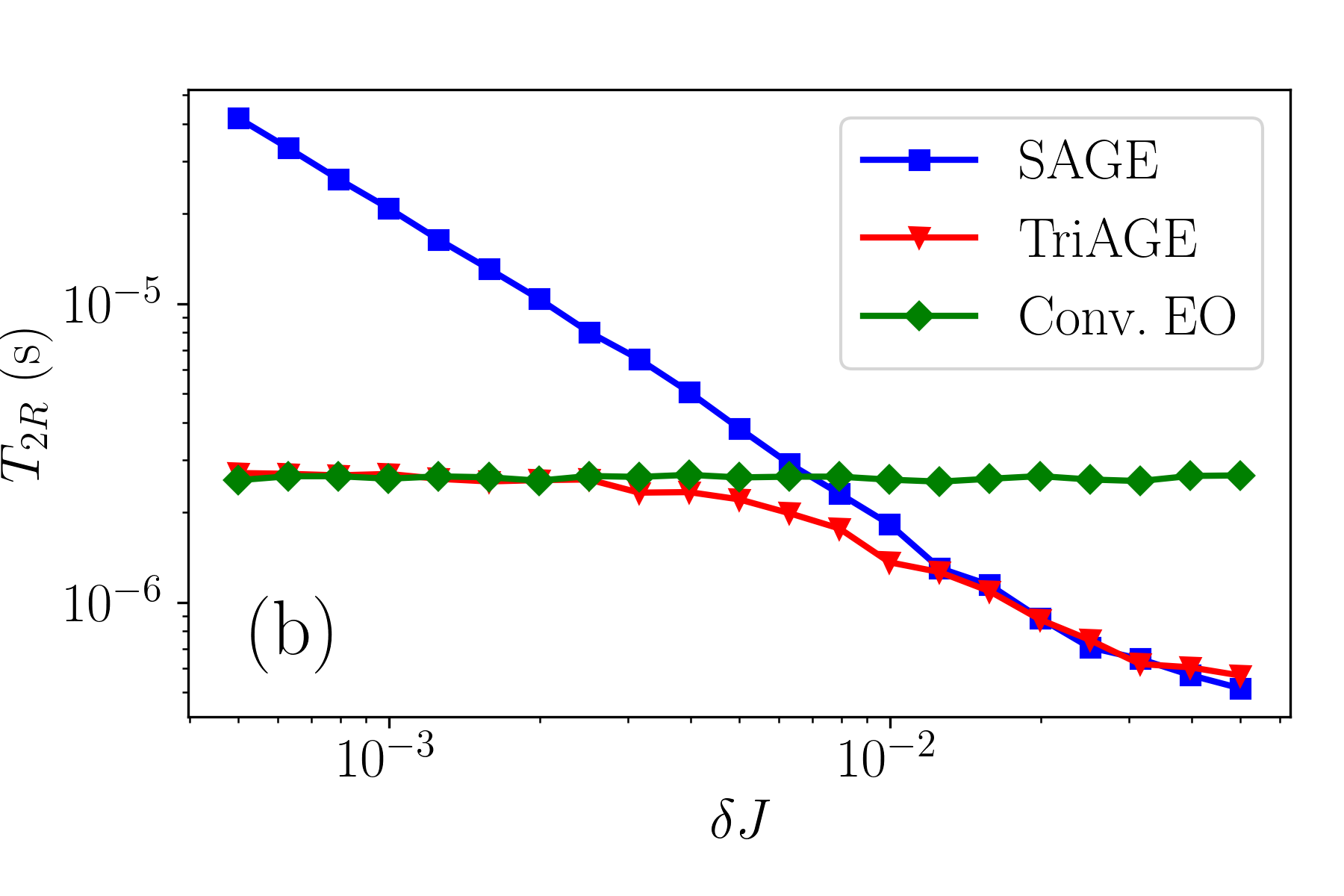}
    \caption{{\it Qubit idle coherence times.} Average idle coherence time for several gapless exchange qubits (see main text for the definitions of the acronyms). 
 For the SAGE and TriAGE qubits, each exchange coupling is set to 10~MHz, whereas all of the conventional EO couplings are set to zero. These decays are averaged over 2500 realizations of magnetic and exchange disorder. (a) Coherence times for a range of magnetic disorder strengths. Each time is extracted from a Gaussian fit of the average idle decay with $\delta J = 5\times10^{-3}$. (b) Coherence times for a range of relative exchange disorders with $\delta h = 50$~kHz. Both $\delta J = 5\times10^{-3}$ and $\delta h = 50$~kHz reflect parameters that may be experimentally attainable. $H_C = 0$ for conventional EO systems, since $J=0$ for each coupling and the noise is multiplicative.}
\label{fig:coherence_times}
\end{figure}

\begin{figure*}[htpb]
    \centering
    \includegraphics[width=0.3\textwidth]{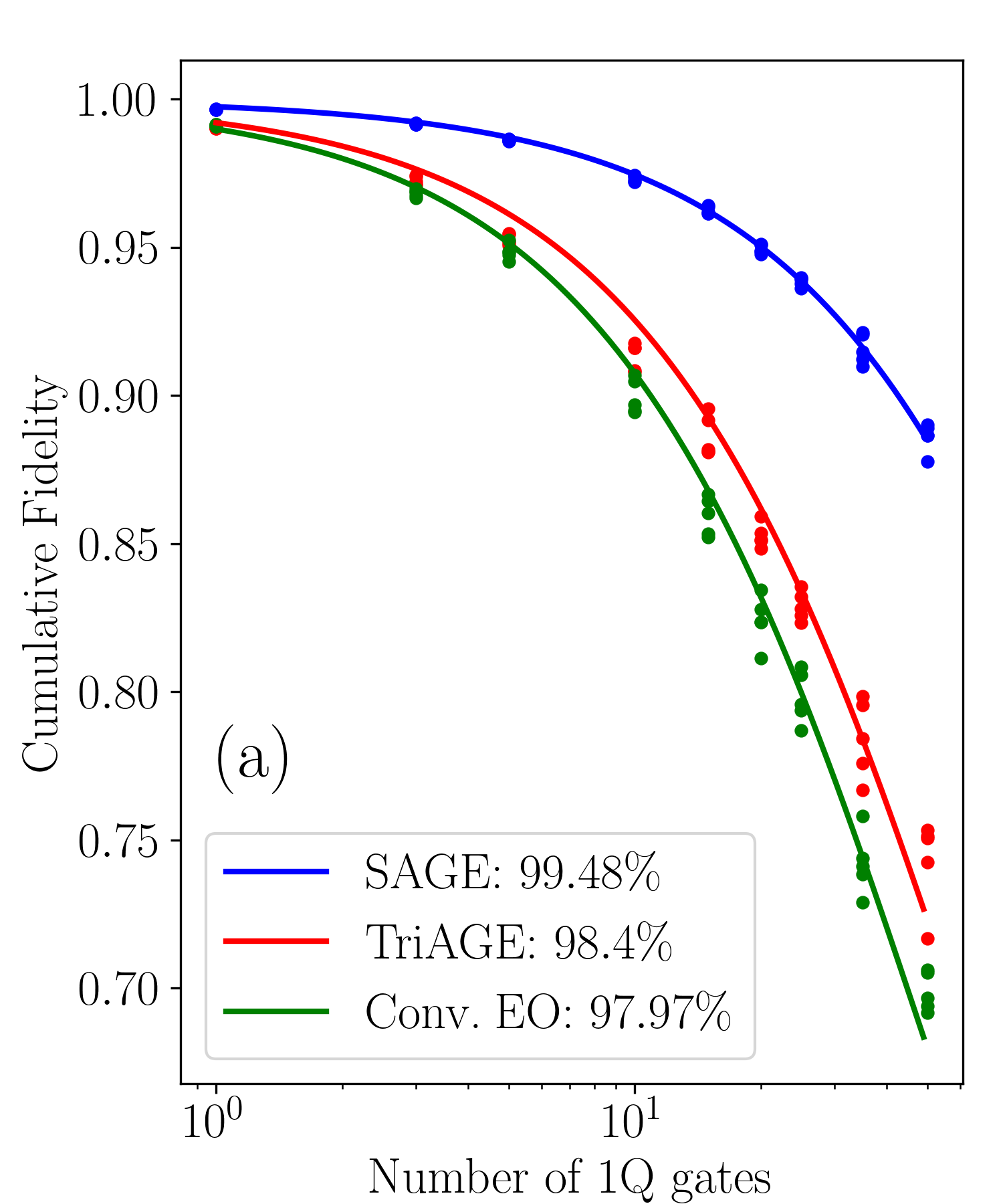}
    \includegraphics[width=0.3\textwidth]{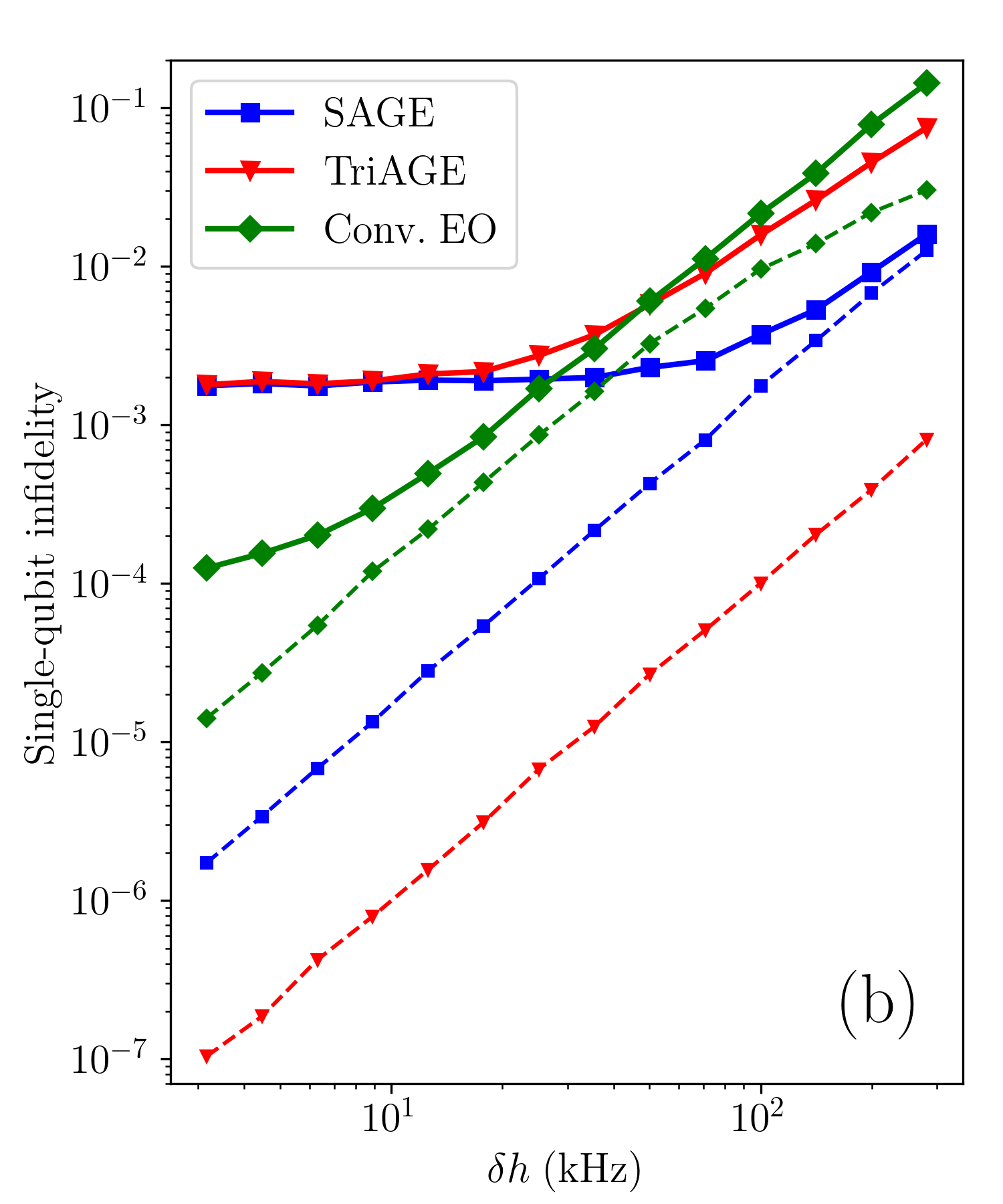}
    \includegraphics[width=0.3\textwidth]{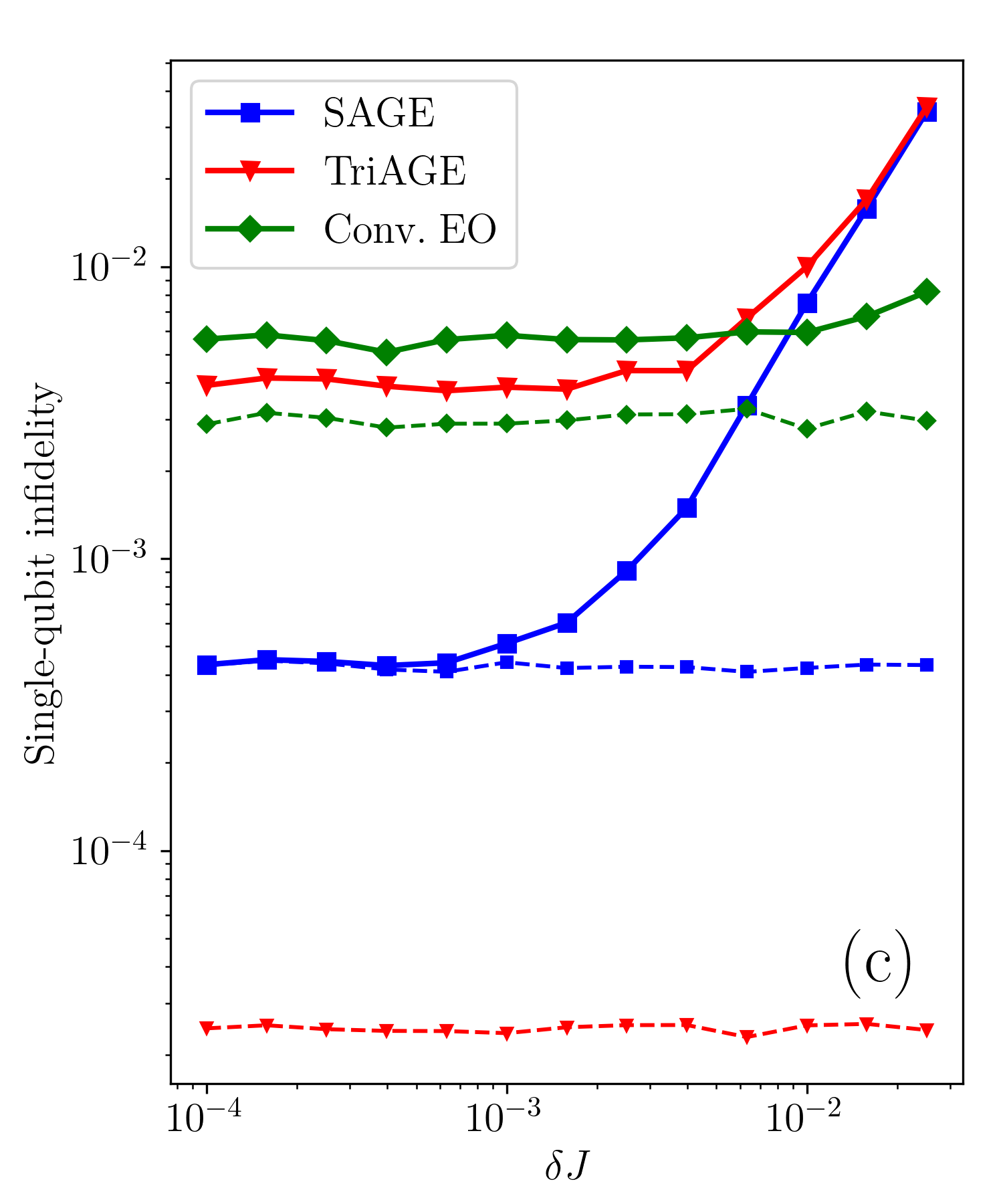}
    \caption{{\it Single-qubit gate fidelities.} (a) Randomized benchmarking results for several gapless exchange qubits. Each point represents an average of 100 gate sequences of a given length, with each sequence averaged over 100 realizations of magnetic and exchange disorder with $\delta h = 100$~kHz and $\delta J = 7\times10^{-3}$, corresponding to $Q^\textrm{eff} \approx 50$ and $T_2^\textrm{eff} \approx 1.3$~$\mu$s. The lines are exponential fits. (b) Fidelities extracted from randomized benchmarking over a sweep of $\delta h$ values, where $\delta J = 5\times10^{-3}$. (c) Fidelities extracted from randomized benchmarking over a sweep of $\delta J$ values, where $\delta h = 50$~kHz. The dashed lines in (b) and (c) represent qubit leakage due to the magnetic gradient $\delta h$.}
\label{fig:fidelities}
\end{figure*}

The primary sources of decoherence in exchange-controlled qubits are magnetic field gradients and charge fluctuations. The fluctuations in magnetic field gradients are modeled by a random magnetic field along the axis of quantization $H_B=\sum_i h_i s^z_i$~\footnote{We assume that a large global magnetic field is applied to the system, which justifies the polarization of magnetic disorder along the axis of quantization.} and fluctuations in charge are modeled by fluctuations in the exchange interaction between the localized electrons $H_C =\frac{1}{4} \sum_{i<j}  J_{ij} \epsilon_{ij} \boldsymbol{s}_i\cdot\boldsymbol{s}_j$ with $i,j\in\{1,2,3,4\}$. These fluctuations are added to the underlying exchange Hamiltonian, resulting in  
$H = H_\textrm{exch} + H_B + H_C$, where $J_{ij}$ (contained in  $H_\textrm{exch}$ and $H_C$) is only nonzero if spins $i$ and $j$ are nearest neighbors in Fig.~\ref{fig:1Q_layout_blochsphere}(a). Upon initializing the encoded qubit along the $x$ axis of the Bloch sphere, the state evolves according to a realization of the noise profile generated by sampling $h_i$ ($\epsilon_{ij}$) from a uniform distribution $[-\delta h,\delta h]$ ($[-\delta J,\delta J]$). The scale of magnetic disorder $\delta h$ is assumed to be dependent on the underlying microscopics of the system and constant for a given realization. The quasistatic charge disorder $\delta J$, however, scales approximately linearly with the exchange coupling strength~\cite{reed_reduced_2016,shulman_demonstration_2012}. Therefore, we vary the strength of disorder in our simulations by sweeping the values of unitful magnetic disorder $\delta h$ and dimensionless charge disorder $\delta J$. The time-dependent off-diagonal element of the qubit density matrix is then averaged over 2500 realizations of noise and fit to $A \exp[-(t/T_{2R})^2]+B$, where $T_{2R}$ is the Ramsey coherence time. To compare coherence times between exchange qubits, we also simulate and subsequently extract the coherence times for conventional EO qubits, composed of three electrons in a linear array, and the triangular always-on gapless exchange-only (TriAGE) analog to SAGE~\cite{acuna_coherent_2024,weinstein_energetic_2005}, which is composed of three electrons in a triangular geometry.

We extract coherence times by first varying magnetic disorder while fixing $\delta J = 5 \times 10^{-3}$, then varying charge disorder while fixing $\delta h = 50$~kHz. 
Averaging over the disordered ensemble, both $\delta J$ and $\delta h$ can be mapped to experimentally relevant parameters: the number of exchange oscillations observed before the qubit decoheres, $Q^\textrm{eff}$, and the dephasing time of a singlet-triplet qubit due to magnetic fluctuations, $T_2^\textrm{eff}$.
Using this approach, we find $\delta J = 5 \times 10^{-3}$ corresponds to $Q^\textrm{eff} \approx 70$ and $\delta h = 50$~kHz corresponds to $T_2^\textrm{eff} \approx 2.6$~$\mu$s.
For both the SAGE and TriAGE qubits, we fix $J_0=10$~MHz, and for conventional EO qubits, the exchange is zero.

For small $\delta h\lesssim30$~kHz [see Fig.~\ref{fig:coherence_times}(a)], the conventional EO qubit has a longer coherence time than both the SAGE and TriAGE qubits whose coherence times are roughly equal and constant. When the magnetic disorder is larger, $\delta h\gtrsim40$~kHz, the SAGE qubit has the longest coherence time and is roughly constant as a function of $\delta h$ while the conventional EO and TriAGE qubits have roughly equal coherence times, which decrease linearly with $\delta h$. This behavior is a consequence of the three-dot qubits being magnetic-disorder-dominated for large values of $\delta h$, whereas the SAGE qubit is limited by charge noise throughout Fig.~\ref{fig:coherence_times}(a). 

Similarly, when fixing $\delta h=50$~kHz and varying $\delta J$ [see Fig.~\ref{fig:coherence_times}(b)], the SAGE qubit coherence time linearly decreases as $\delta J$ is increased and has the best coherence times relative to the others when $\delta J\lesssim 7 \times 10^{-3}$. The conventional EO qubit is unaffected by fluctuations in charge noise since the exchange couplings vanish during a conventional EO idle, but the coherence time monotonically increases as the variance in magnetic gradient decreases. Because the SAGE qubit is protected from magnetic noise when $J_0\gg\delta h$, 
the decoherence is dominated by charge noise. The TriAGE qubit is susceptible to both coherent magnetic gradient variance and charge noise, such that it inherits poor coherence times over all values of $\delta h$.

\paragraph{Single-qubit operation.}
We characterize single-qubit gate fidelity using randomized benchmarking~\cite{knill_randomized_2008}. We apply $N_\textrm{1Q}$ gates sampled from a set of single-qubit Cliffords (see SM~\cite{supplemental}) for a given disorder realization and average the fidelity over $100$ disorder realizations and $100$ random gate sequences to obtain a cumulative fidelity [Fig.~\ref{fig:fidelities}(a)]. For each single-qubit gate, we set the largest exchange coupling $J_0 = 10~$MHz. Our error channel, although unitary and quasistatic, is depolarized using the entire Clifford gate set, resulting in a pure exponential decay rate of the cumulative fidelity~\cite{sheldon_characterizing_2016} as a function of the sequence length $N_\textrm{1Q}$.

We extract single-qubit gate fidelity as a function of $\delta h$ while fixing $\delta J$ [Fig.~\ref{fig:fidelities}(b)].
When $\delta h\lesssim30~$kHz, the SAGE and TriAGE single-qubit gates are charge-noise-limited and have a worse fidelity than the conventional EO single-qubit gates, given $\delta J=5 \times 10^{-3}$. For larger values of $\delta h$, the single-qubit gate infidelities are significantly improved for SAGE compared to the TriAGE and conventional EO qubit fidelities. Fig.~\ref{fig:fidelities}(c) illustrates single-qubit gate performance for various values of $\delta J$. For $\delta h = 50$~kHz, the SAGE qubit outperforms the two three-dot exchange qubits up to approximately $\delta J = 9 \times 10^{-3}$. The conventional EO qubit fidelity is nearly constant with respect to charge noise, as seen in Fig.~\ref{fig:fidelities}(c) due to the magnetic fluctuations dominating the error in this regime. Although TriAGE qubits have lower fidelity than SAGE or conventional EO for any disorder value, they demonstrate enhanced leakage suppression due to the large gap between leakage and computational states.

The SAGE single-qubit fidelity in Fig.~\ref{fig:fidelities}(b) becomes affected by magnetic disorder beyond $\delta h = 70$~kHz. Using the same parameters for idle coherence times in Fig.~\ref{fig:coherence_times}(a), this value was approximately $\delta h = 300$~kHz. Thus, we see that the SAGE qubit is especially resilient to magnetic disorder when idling since the exchange couplings are maximal in this case. When performing gates, exchange couplings are lowered, thereby weakening the energetic leakage suppression.

\begin{figure*}[htbp]
  \includegraphics[width=0.9\textwidth]{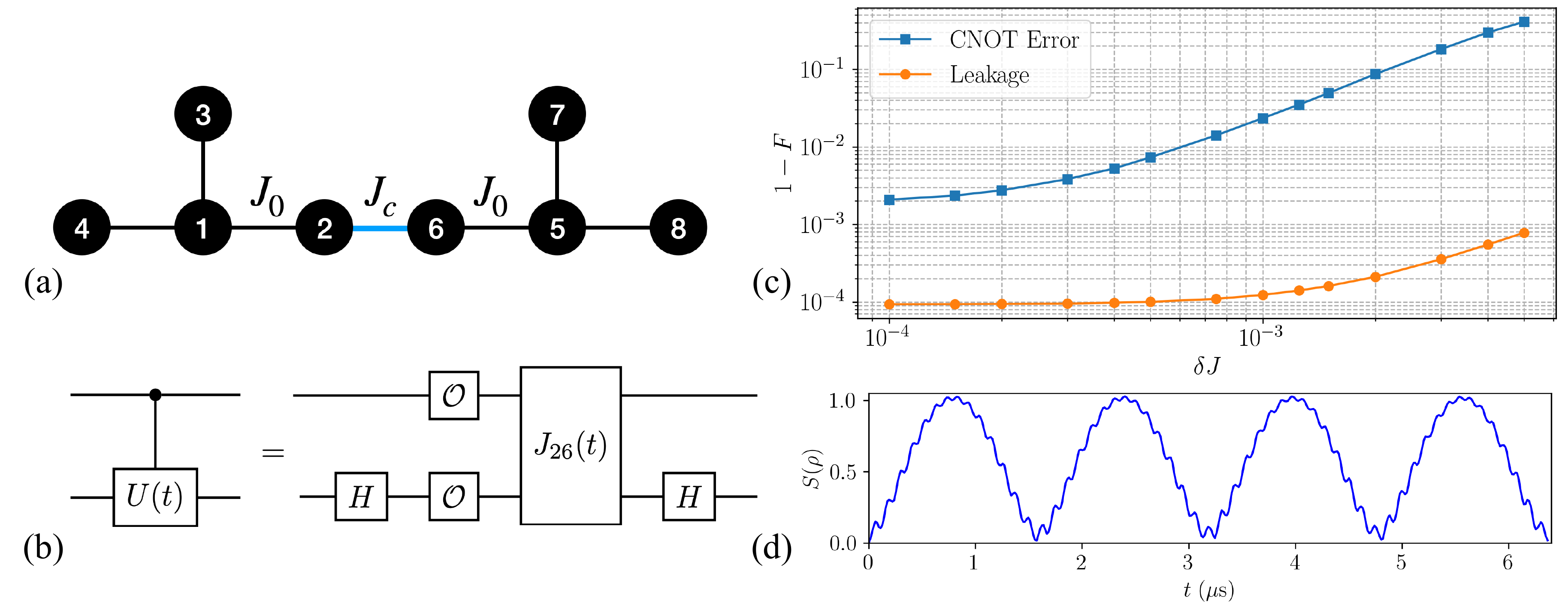}

  \caption{\textit{Two-qubit gates.} (a) Example geometry to perform a two-qubit gate using a single interqubit exchange pulse $J_c = J_{26}$. (b) A single $J_{26}(t)$ pulse up to local unitaries is equivalent to a controlled-$\mathcal{U}(t)$ rotation. $\mathcal{U}(t_\textrm{CNOT}) = X$ and the two-qubit circuit represents a CNOT gate. $H$ is the Hadamard gate and $\mathcal{O}$ is $\exp(\frac{i\pi \theta}{2}\sigma_z)$ where $\theta \approx \tfrac{3}{8} + \frac{9J_c}{128J_0}$. Each gate can be performed in a single pulse. (c) The average CNOT infidelity and leakage probability for a SAGE two-qubit gate experiencing quasistatic noise (constant for the gate duration), where $\delta h = 50$~kHz. (d) A representative simulation of the von Neumann entanglement entropy of the initially separable state $|\psi_0\rangle = \frac{1}{\sqrt{2}} \bigl(|00\rangle + |10\rangle\bigr)$ as a result of the SAGE two-qubit interaction with $J_c = 4$~MHz and $J_0 = 20$~MHz to facilitate a CNOT time $t_\textrm{CNOT} \approx \ 785$~ns.}
  \label{fig:2q_layout_example}
\end{figure*}

\paragraph{Two-qubit operation.} 
SAGE two-qubit gates can be performed with a single interqubit exchange pulse $J_c$. Although, in general, turning on an interqubit exchange interaction results in leakage to noncomputational states, this is suppressed by the always-on intraqubit exchange interactions. Therefore, to have good fidelity, we must operate in the regime $J_c\ll J_0$. In this limit, we can perturbatively extract the effective interqubit interaction using a Schrieffer-Wolff transformation, which projects the full 8-dot Hilbert space dynamics onto the four-dimensional computational subspace of the two qubits. A relatively simple effective two-qubit interaction
\begin{equation} 
H_\textrm{2Q}^\textrm{eff} = \left(\frac{J_c^2}{6J_0} + \frac{5J_c^3}{32J_0^2}\right) \sigma^z_1\sigma^z_2 - \left(\frac{J_c^2}{24J_0} + \frac{J_c^3}{64J_0^2}\right)\sigma^z_A ,
\label{eq:main_schriefferwolff}
\end{equation} 
where $\sigma_A^z = (\sigma_1^z + \sigma_2^z)$ and $\sigma_j^z$ is the Pauli matrix acting on the $j$th qubit, is achieved by pulsing the exchange interaction between each qubit's second electron [the electron that forms an encoding pair with the qubit's core electron---labeled 2 and 6 in Fig.~\ref{fig:2q_layout_example}(a)]. Because $J_c$ must be relatively large in order to have a fast gate, we calculated Eq.~(\ref{eq:main_schriefferwolff}) up to third order in $J_c/J_0$ (see SM~\cite{supplemental}).

We also simulate this interaction exactly, i.e. using all computational and noncomputational states, using $J_0=20$~MHz and $J_c=4$~MHz. We find that the resulting gate is locally equivalent to a CNOT, and the intrinsic fidelity of this gate (i.e. in the absence of disorder) is $\sim99.8$\% ($99.99$\% free of leakage) with a gate time of approximately $785$~ns. Quasistatic charge noise over the entire life of the operation reduces the gate's fidelity while mostly preserving its prevention of leakage [see Fig.~\ref{fig:2q_layout_example}(c)]. The circuit diagram for a CNOT gate is shown in Fig.~\ref{fig:2q_layout_example}(b). In all, one interqubit and four intraqubit gate operations are needed to achieve a CNOT. To demonstrate that this interaction is indeed maximally entangling, we simulate its effect on the entanglement entropy of an initially separable state $|\psi_0\rangle = \frac{1}{\sqrt{2}} \bigl(|00\rangle + |10\rangle\bigr)$ in Fig.~\ref{fig:2q_layout_example}(d) as a function of the duration $t$ of the $J_{c}$ pulse.

Due to its adiabatic nature, the SAGE two-qubit gate is fidelity- and speed-limited by the ratio $J_c/J_0$. In particular, we numerically extract these functional dependencies (see SM~\cite{supplemental}) for the lower bound on the intrinsic fidelity $F_\textrm{CNOT}$ with a gate duration of $t_\textrm{CNOT}$:
\begin{equation}\label{eq:fcnot}
 F_\textrm{CNOT} \approx 1 - \frac{J_c^2}{2 J_0^2}\,, \quad\quad t_\textrm{CNOT} \approx \frac{3J_0} {4J_c^2} - \frac{45}{64J_c}\,.
\end{equation}
Intrinsic gate fidelities can be improved beyond this lower bound through careful selection of $J_c$ and $J_0$ so that leakage fluctuations [as seen in Fig.~\ref{fig:2q_layout_example}(d)] are not the principal source of error. Indeed, our example with $J_0=20$~MHz and $J_c=4$~MHz shown in Fig.~\ref{fig:2q_layout_example}(d)  corresponds to such a favorable case with a fidelity far above the lower bound. Furthermore, increasing the maximal intraqubit exchange interaction enables faster and higher fidelity CNOT gates. For example, an intraqubit exchange interaction of $J_0=100$~MHz would allow a commensurately larger $J_c$ of 20~MHz, enabling a gate time $\sim 157$~ns while maintaining $\sim99.8$\% fidelity.

\paragraph{Discussion.}

The SAGE architecture's resilience to differences in Zeeman splitting makes it similarly immune to variations in the $g$-factor between dots. Consequently, while conventional EO qubits suffer from poor qubit fidelities as the magnetic field is increased beyond $\sim1$~mT~\cite{weinstein_universal_2023}, larger global magnetic fields can be applied to SAGE qubits. This insensitivity to $g$-factor variations may also allow for realizing a SAGE qubit using hole spins if there exists a suitable encoding that can overcome Dzyaloshinskii-Moriya interactions induced by noncolinear $g$-factors and strong spin-orbit interactions~\cite{scappucciNATRM21}.

The always-on gapless encoding enables both clockwise and counterclockwise rotations about any vector in the $x$-$z$ plane through simultaneous operation of two exchange interactions. For example, an $X$ gate can be implemented by setting $J_{14}=J_{12}+\Delta$ and $J_{13}=J_{12}-\Delta$. Any single-qubit rotation can therefore be performed using two pulses.

Two-qubit gate times are largely limited by the intraqubit exchange interaction. Large, controllable exchange interactions have been demonstrated experimentally~\cite{weinstein_universal_2023}. A full configuration interaction calculation and realistic simulation of the electrostatic environment, i.e. taking into account gates, SETs, and electron reservoirs in addition to relative dot positions~\cite{acuna_coherent_2024, george_12-spin-qubit_2024}, would be useful to determine the maximal value that the always-on exchange interaction can take. Moreover, such a calculation would inform the most effective dot geometry for the above T-shaped connectivity.

We have shown that SAGE spin qubits can exhibit improved idle coherence times and higher single-qubit gate fidelities than similar exchange-only qubit designs in experimentally relevant noise regimes. We have outlined a scheme for performing two-qubit gates, including the procedure for CNOT, specifically using a single interqubit exchange pulse and local unitaries. We also verified its efficacy, showing that this gate can be performed in a reasonable time.

The SAGE qubit offers a promising approach to the principal challenges facing the spin qubit community when it comes to scaling up, such as unavoidable magnetic field fluctuations, the heating effects of ac driving, and the impracticalities resulting from on-chip micromagnets.

\begin{acknowledgments}
\paragraph{Acknowledgments.} We thank N. Bishop, F. Mohiyaddin, and M. Curry for helpful discussions about experimental considerations. SH would like to thank A. Mills for valuable discussions. This work was supported by the Laboratory for Physical Sciences.

\end{acknowledgments}
\bibliography{references4}

\clearpage
\newpage
\mbox{~}
\onecolumngrid

\begin{center}
  \textbf{\large {Supplemental Material: \\ Singlet-only Always-on Gapless Exchange Qubits with Baseband Control
 }\\[.2cm]}
\end{center}

\setcounter{equation}{0}
\setcounter{figure}{0}
\setcounter{table}{0}
\setcounter{section}{0}

\renewcommand{\theequation}{S\arabic{equation}}
\renewcommand{\thefigure}{S\arabic{figure}}
\renewcommand{\thesection}{S\arabic{section}}
\renewcommand{\thetable}{S\arabic{table}}
\renewcommand{\bibnumfmt}[1]{[S#1]}
\renewcommand{\citenumfont}[1]{S#1}
\section{Single-qubit Hamiltonians}
\subsection{Four-electron Hamiltonian}

The disordered Heisenberg Hamiltonian with an onsite Zeeman term,
\begin{equation}
 H = \frac{1}{4}\sum_{\langle i,j\rangle} J_{ij} \boldsymbol{s}_i\cdot\boldsymbol{s}_j - \sum_i h_i s^z_i,
    \label{eq:simulation_ham}
\end{equation}
restricts single-qubit dynamics to a certain $S_z$ subspace. For the SAGE qubit, we work in the $S_z=0$ subspace of a four-electron system. This subspace is spanned by six states $\ket{0}$, $\ket{1}$,
$\ket{T_1}$, $\ket{T_2}$, $\ket{T_3}$, and $\ket{Q}$ that can be characterized by the quantum numbers $S$, $ S_{12}$, and $S_{34}$, where $S_{12}$ ($S_{34}$) is the total spin angular momentum eigenvalue for the electron pair \{1,2\} (\{3,4\}):
\[
\begin{tabular}{ c|c c c c c c}
     & \ket{0} & \ket{1} & 
    \ket{T_1} & \ket{T_2} & \ket{T_3} & \ket{Q} \\ \hline
     $S$ & 0 & 0 & 1 & 1 & 1 & 2 \\
     $S_{12}$ & 0 & 1 & 0 & 1 & 1 & 1 \\
     $S_{34}$ & 0 & 1 & 1 & 0 & 1 & 1\\
     $S_z$ & 0 & 0 & 0 & 0 & 0 & 0\\
\end{tabular}
\begin{aligned}
\hspace{1cm}&\lvert0\rangle = \frac{1}{2}\bigl(\lvert\uparrow\downarrow\uparrow\downarrow\rangle - \lvert\uparrow\downarrow\downarrow\uparrow\rangle - \lvert\downarrow\uparrow\uparrow\downarrow\rangle + \lvert\downarrow\uparrow\downarrow\uparrow\rangle\bigr),\\
&\lvert1\rangle = \frac{1}{\sqrt{3}}\Bigl(\lvert\uparrow\uparrow\downarrow\downarrow\rangle 
 - \tfrac12\bigl(\lvert\uparrow\downarrow\uparrow\downarrow\rangle+\lvert\uparrow\downarrow\downarrow\uparrow\rangle+\lvert\downarrow\uparrow\uparrow\downarrow\rangle+\lvert\downarrow\uparrow\downarrow\uparrow\rangle\bigr)
 + \lvert\downarrow\downarrow\uparrow\uparrow\rangle\Bigr),\\
&\lvert T_1\rangle = \frac{1}{2}\bigl(\lvert\uparrow\downarrow\uparrow\downarrow\rangle + \lvert\uparrow\downarrow\downarrow\uparrow\rangle - \lvert\downarrow\uparrow\uparrow\downarrow\rangle - \lvert\downarrow\uparrow\downarrow\uparrow\rangle\bigr),\\
&\lvert T_2\rangle = \frac{1}{2}\bigl(\lvert\uparrow\downarrow\uparrow\downarrow\rangle - \lvert\uparrow\downarrow\downarrow\uparrow\rangle + \lvert\downarrow\uparrow\uparrow\downarrow\rangle - \lvert\downarrow\uparrow\downarrow\uparrow\rangle\bigr),\\
&\lvert T_3\rangle = \frac{1}{\sqrt{2}}\bigl(\lvert\uparrow\uparrow\downarrow\downarrow\rangle - \lvert\downarrow\downarrow\uparrow\uparrow\rangle\bigr),\\
&\lvert Q\rangle = \frac{1}{\sqrt{6}}\bigl(\lvert\uparrow\uparrow\downarrow\downarrow\rangle + \lvert\uparrow\downarrow\uparrow\downarrow\rangle + \lvert\uparrow\downarrow\downarrow\uparrow\rangle + \lvert\downarrow\uparrow\uparrow\downarrow\rangle + \lvert\downarrow\uparrow\downarrow\uparrow\rangle + \lvert\downarrow\downarrow\uparrow\uparrow\rangle\bigr).
\end{aligned}
\]

In this basis, $H$ (restricted to the $S_z=0$ subspace) has the matrix representation
\begin{equation*}
 H^\textrm{4-dot}_\textrm{1Q}
 =
  \begin{pmatrix}
 -\tfrac{3}{4}J_a
    & \tfrac{\sqrt{3}}{4}(J_c - J_b)
    & -\Delta_{34}
    & -\Delta_{12}
    & 0
    & 0
    \\[6pt]
 \tfrac{\sqrt{3}}{4}(J_c - J_b)
    & \tfrac{1}{4}\bigl(J_a - 2(J_b + J_c)\bigr)
    & \tfrac{1}{\sqrt{3}}\Delta_{12}
    & \tfrac{1}{\sqrt{3}}\Delta_{34}
    & \sqrt{\tfrac{2}{3}}\bigl(-\Delta_{13}-\Delta_{24}\bigr)
    & 0
    \\[6pt]
 -\Delta_{34}
    & \tfrac{1}{\sqrt{3}}\Delta_{12}
    & -\tfrac{1}{4}\bigl(J_a + 2J_{\Delta a}\bigr)
    & \tfrac{1}{4}(J_b - J_c)
    & -\tfrac{\sqrt{2}}{4}(J_{\Delta b} + J_{\Delta c})
    & -\sqrt{\tfrac{2}{3}}\Delta_{12}
    \\[6pt]
 -\Delta_{12}
    & \tfrac{1}{\sqrt{3}}\Delta_{34}
    & \tfrac{1}{4}(J_b - J_c)
    & \tfrac{1}{4}\bigl(2J_{\Delta a} - J_a\bigr)
    & \tfrac{\sqrt{2}}{4}(J_{\Delta b} - J_{\Delta c})
    & -\sqrt{\tfrac{2}{3}}\Delta_{34}
    \\[6pt]
    0
    & \sqrt{\tfrac{2}{3}}\bigl(-\Delta_{13}-\Delta_{24}\bigr)
    & -\tfrac{\sqrt{2}}{4}(J_{\Delta b} + J_{\Delta c})
    & \tfrac{\sqrt{2}}{4}(J_{\Delta b} - J_{\Delta c})
    & \tfrac{1}{4}(J_a - J_b - J_c)
    & \tfrac{-\Delta_{13}-\Delta_{24}}{\sqrt{3}}
    \\[6pt]
    0
    & 0
    & -\sqrt{\tfrac{2}{3}}\Delta_{12}
    & -\sqrt{\tfrac{2}{3}}\Delta_{34}
    & \tfrac{-\Delta_{13}-\Delta_{24}}{\sqrt{3}}
    & \tfrac{1}{4}(J_a + J_b + J_c)
  \end{pmatrix},
\end{equation*} 
where $\Delta_{ij} = h_i - h_j$, $J_a = J_{12}+J_{34}$, $J_b = J_{13} + J_{24}$, $J_c = J_{14} + J_{23}$, $J_{\Delta a} = J_{12} - J_{34}$, $J_{\Delta b} = J_{13} - J_{24}$, and $J_{\Delta c} = J_{14} - J_{23}$. When setting $J_{23}=J_{24}=J_{34}=0$ in the absence of magnetic disorder, we obtain the single-qubit Hamiltonian given in Eq.~(3) of the main text up to a constant shift.

$H^\textrm{4-dot}_\textrm{1Q}$ makes it visually clear that magnetic field gradients only couple states that have different values of $S$. 
To ensure optimal energetic suppression of leakage, the energy gaps between the computational states with $S=0$ and the leakage states with $S=1,2$ should be maximized. At the same time, the computational states should remain degenerate to allow for gapless operation of the qubit. 

Diagonalizing $H_\textrm{1Q}^\textrm{4-dot}$ in the absence of magnetic disorder, we find that gapless operation requires $J_a = J_b = J_c = J$, so that both qubit states have energy $-\frac{3}{4}J$. 
Furthermore, the gaps between computational and leakage states are found to reach a maximal value of $\frac{1}{2}J$ at two optimal points in parameter space. The first optimal point is $J_{\Delta a} = J_{\Delta b} = J_{\Delta c} = \pm J$, which implies a T-shape geometry, as outlined in the main text. The second optimal point is $J_{\Delta a} = J_{\Delta b} = J_{\Delta c} = 0$, which implies a box-shape geometry with all-to-all connectivity. This second geometry has been explored theoretically~\cite{bacon_coherence-preserving_2001,weinstein_quantum-dot_2005,weinstein_scalable_2007,antonio_two-qubit_2013,jiang_preparation_2009} but would be significantly more difficult to implement in quantum dots than the simple T-shape we propose. Other simple connectivities such as a linear array or a square loop are incompatible with the $J_a = J_b = J_c$ requirement, and therefore cannot be used for SAGE qubits. Additionally, these connectivities have leakage ``hotspots'' where, for certain values of the nearest-neighbor exchange couplings, a leakage state is degenerate with the computational states. This is not the case for the T-shape geometry, where the leakage states are always separated from the computational states if all three exchange couplings are nonzero.

\begin{figure}

    \begin{minipage}{0.4\linewidth} 
    \begin{tikzpicture}
        \draw[black, line width=0.5mm] (1,3) -- (2.5,3) node[below left, xshift=-1.1mm]{};
        \draw[black, line width=0.5mm] (2.5,3) -- (4,3) node[below left, xshift=-1.1mm]{};
        \draw[black, line width=0.5mm] (2.5,3) -- (2.5,4.5) node[below right, yshift=-2.5mm]{};
        \node[circle, draw, fill=black, text centered, minimum size=5mm] at (2.5, 3) {};
        \node[circle, draw, fill=black, text centered, minimum size=5mm] at (1, 3) {};
        \node[circle, draw, fill=black, text centered, minimum size=5mm] at (2.5, 4.5) {};
        \node[circle, draw, fill=black, text centered, minimum size=5mm] at (4, 3) {};
    \end{tikzpicture}
    \end{minipage}
    \hspace{-20mm}
    \begin{minipage}{0.4\linewidth} 
    \begin{tikzpicture}
        
        \draw[black, line width=0.5mm] (0,0) -- (0,2) node[below left, xshift=-1.1mm]{};
        \draw[black, line width=0.5mm] (0,0) -- (2,0) node[below left, xshift=-1.1mm]{};
        \draw[black, line width=0.5mm] (0,0) -- (2,2) node[below left, xshift=-1.1mm]{};
        \draw[black, line width=0.5mm] (0,2) -- (2,0) node[below left, xshift=-1.1mm]{};
        \draw[black, line width=0.5mm] (0,2) -- (2,2) node[below left, xshift=-1.1mm]{};
        \draw[black, line width=0.5mm] (2,0) -- (2,2) node[below left, xshift=-1.1mm]{};

        \node[circle, draw, fill=black, text centered, minimum size=5mm] at (0, 2) {};
        \node[circle, draw, fill=black, text centered, minimum size=5mm] at (0, 0) {};
        \node[circle, draw, fill=black, text centered, minimum size=5mm] at (2, 2) {};
        \node[circle, draw, fill=black, text centered, minimum size=5mm] at (2, 0) {};
    \end{tikzpicture}
    \end{minipage}
    \vspace{2mm}
    \caption{The two optimal gapless qubit connectivities to ensure maximal leakage suppression. Other geometries are possible for four-electron always-on qubits, but they are either not gapless---such as a linear array---or suffer from enhanced leakage compared to these two geometries.}
\end{figure}

\subsection{Three-electron Hamiltonian}

For the three-electron encodings discussed in the main text, i.e., the conventional EO qubit and the always-on gapless exchange qubit (TriAGE), we work in the total $S_z=1/2$ subspace of a three-electron system. This subspace is spanned by three states $|S,S_{12}\rangle = \{|\frac{1}{2},0\rangle,|\frac{1}{2},1\rangle, |\frac{3}{2},1\rangle\}$. The first two states are the computational states, while the third state is a leakage state. The states are given by
\begin{align*}
|0\rangle &= |\frac{1}{2},0\rangle = |S \rangle|\!\uparrow\rangle,\\
|1\rangle &= |\frac{1}{2},1\rangle = \frac{1}{\sqrt{3}}\Bigl(\sqrt{2}|T_+\rangle|\!\downarrow\rangle - |T_0 \rangle|\!\uparrow\rangle \Bigr),\\
|L\rangle &= |\frac{3}{2},1\rangle = \frac{1}{\sqrt{3}}\Bigl(|T_+\rangle|\!\downarrow\rangle + \sqrt{2}|T_0 \rangle|\!\uparrow\rangle \Bigr),
\end{align*}
where $\ket{S} = (\ket{\uparrow\downarrow} - \ket{\downarrow\uparrow})/\sqrt{2}$,  $\ket{T_0} = (\ket{\uparrow\downarrow} + \ket{\downarrow\uparrow})/\sqrt{2}$, and $\ket{T_+} = \ket{\uparrow\uparrow}$.
The Hamiltonian defined in Eq.~(\ref{eq:simulation_ham}), restricted to the total $S_z={1/2}$ subspace, can then be expressed in matrix form (up to a constant energy offset) as:
\begin{equation*}
H_{\textrm{1Q}}^{\textrm{3-dot}}
= \frac{1}{3}
\begin{pmatrix}
\Delta_{13} + \Delta_{23} - \tfrac{9}{4}J_{12}
&
\sqrt{3}\,\bigl(\Delta_{12} - \tfrac{3}{4}J_{13} + \tfrac{3}{4}J_{23}\bigr)
&
-\sqrt{6}\,\Delta_{12}
\\[6pt]
\sqrt{3}\,\bigl(\Delta_{12} - \tfrac{3}{4}J_{13} + \tfrac{3}{4}J_{23}\bigr)
&
-\Delta_{13} - \Delta_{23} + \tfrac{3}{4}J_{12} - \tfrac{3}{2}\,(J_{13} + J_{23})
&
-\sqrt{2}\,(\Delta_{13} + \Delta_{23})
\\[6pt]
-\sqrt{6}\,\Delta_{12}
&
-\sqrt{2}\,(\Delta_{13} + \Delta_{23})
&
\tfrac{3}{4}\,(J_{12} + J_{13} + J_{23})
\end{pmatrix}
\end{equation*}
where $\Delta_{ij} = h_i - h_j$. 
Unlike the four-dot case, $H_{\textrm{1Q}}^\textrm{3-dot}$ contains magnetic-gradient-induced couplings between computational states, making the energetic suppression of magnetic-gradient-induced errors impossible.

\section{Single-qubit gates used in randomized benchmarking}

Our randomized benchmarking results in the main text are calculated after initializing the qubit to the state $\ket{\psi} = \frac{1}{\sqrt{2}}\left(\ket{0} + i \ket{1}\right)$ using the complete set of single-qubit Clifford gates, which are listed in Table~\ref{tab:cliffords}. Each Clifford gate is performed using the always-on gapless primitive gates in the case of SAGE and TriAGE. In the case of conventional EO, the Cliffords are performed following the procedure in Ref.~\cite{weinstein_universal_2023}. The exchange strengths and pulse durations for the primitive AGE gates are laid out in Table~\ref{tab:how-to}.
\vspace{3mm}

\begin{table}[ht]
\centering
\[
\begin{array}{|c|c|c|}
\hline
\textbf{Angle} & \textbf{Axis} & \textbf{AGE gates} \\
\hline
- & - & \text{Identity}\\
\hline
\pi & \hat{x} & X \\
\pi & \hat{y} & Z X \\
\pi & \hat{z} & Z \\
\hline
+\frac{\pi}{2} & \hat{z} & S \\
-\frac{\pi}{2} & \hat{z} & S^{\dagger} \\
+\frac{\pi}{2} & \hat{x} & \sqrt{X} \\
-\frac{\pi}{2} & \hat{x} & \sqrt{X}^{\dagger} \\
+\frac{\pi}{2} & \hat{y} & X H \\
-\frac{\pi}{2} & \hat{y} & Z H \\
\hline
\pi & \frac{\hat{x} + \hat{z}}{\sqrt{2}} & H \\
\pi & \frac{\hat{x} - \hat{z}}{\sqrt{2}} & H' \\
\pi & \frac{\hat{x} + \hat{y}}{\sqrt{2}} & X S \\
\pi & \frac{\hat{x} - \hat{y}}{\sqrt{2}} & X S^{\dagger} \\
\pi & \frac{\hat{y} + \hat{z}}{\sqrt{2}} & Z \sqrt{X}^{\dagger} \\
\pi & \frac{\hat{y} - \hat{z}}{\sqrt{2}} & Z \sqrt{X} \\
\hline
+\frac{\pi}{3} & (1, 1, 1) & H S \\
-\frac{\pi}{3} & (1, 1, 1) & H \sqrt{X} \\
+\frac{\pi}{3} & (-1, 1, 1) & S^{\dagger} \sqrt{X} \\
-\frac{\pi}{3} & (-1, 1, 1) & \sqrt{X} S^{\dagger} \\
+\frac{\pi}{3} & (1, -1, 1) & \sqrt{X}^{\dagger} S \\
-\frac{\pi}{3} & (1, -1, 1) & S \sqrt{X}^{\dagger} \\
+\frac{\pi}{3} & (1, 1, -1) & H S^{\dagger} \\
-\frac{\pi}{3} & (1, 1, -1) & H \sqrt{X}^{\dagger} \\
\hline
\end{array}
\]
\caption{Single-qubit Clifford gates used in randomized benchmarking. The AGE gates are the decomposition of each Clifford gate into gates that are primitive for an always-on gapless exchange qubit, allowing each Clifford gate to be performed in at most two pulses. $H'$ represents a Hadamard-like operation that can be performed by AGE qubits in a single exchange pulse.
\label{tab:cliffords}}
\end{table}
\begin{table}[h]
\centering
\begin{tabular}{|c|c|c|c|c|c|c|c|c|c|}
\hline
              & $I$ & $X$ & $Z$ & $S$ & $S^+$ & $\sqrt{X}$ & $\sqrt{X}^+$ & $H$ & $H'$ \\ \hline
$J_{a}$      & $1$ & $1-\tfrac1{2\sqrt3}$ & $\tfrac12$ & $\tfrac12$ & $1$ & $1-\tfrac1{2\sqrt3}$ & $1-\tfrac1{2\sqrt3}$ & $1-\tfrac1{2\sqrt2}-\tfrac1{2\sqrt6}$ & $1-\tfrac1{2\sqrt2}-\tfrac1{2\sqrt6}$ \\ \hline
$J_{b}$      & $1$ & $1-\tfrac1{\sqrt3}$   & $1$        & $1$        & $\tfrac12$ & $1-\tfrac1{\sqrt3}$   & $1$                   & $1-\tfrac1{\sqrt6}$       & $1$                           \\ \hline
$J_{c}$      & $1$ & $1$                   & $1$        & $1$        & $\tfrac12$ & $1$                   & $1-\tfrac1{\sqrt3}$   & $1$                       & $1-\tfrac1{\sqrt6}$           \\ \hline
$\varphi$     & $\tfrac12$ & $\tfrac12$ & $\tfrac12$ & $\tfrac14$ & $\tfrac14$ & $\tfrac14$ & $\tfrac14$ & $\tfrac12$ & $\tfrac12$ \\ \hline
\end{tabular}
\caption{The exchange interaction strengths (arbitrary units) for each AGE primitive gate used to build the Clifford set and its corresponding rotation angle $\varphi$ (in units of $2\pi$). For SAGE, $J_a = J_{12}$, $J_b = J_{13}$, and $J_c = J_{14}$. For TriAGE, $J_a = J_{12}$, $J_b = J_{13}$, and $J_c = J_{23}$.
\label{tab:how-to}}
\end{table}

\section{Characterization of the SAGE two-qubit gate}

\begin{figure}
    \centering
    \hspace{2.8cm}
    \includegraphics[width=0.6\linewidth]{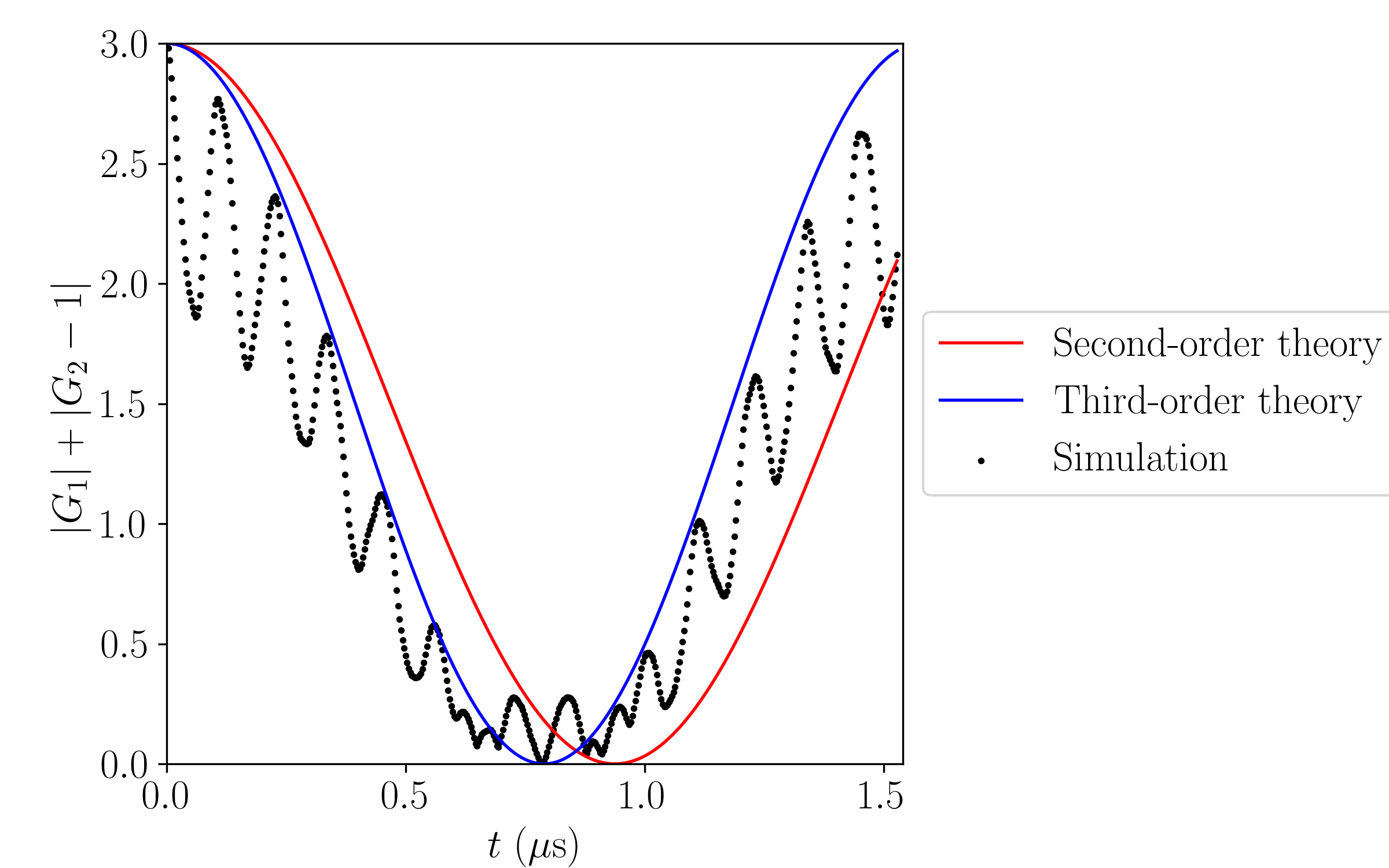}
    \caption{Norm of the deviation of the Makhlin invariants $(G_1,G_2)$ from the correct CNOT invariants. At each time step, the unitary is computed where $J_{26} = 0.2 J_0 = 4$~MHz is the only nonzero interqubit coupling. The Makhlin invariants corresponding to CNOT are $G_1 = 0$ and $G_2 = 1$. }
    \label{fig:SchrifferWolff}
\end{figure}

We perform a Schrieffer-Wolff transformation to understand the effective two-qubit Hamiltonian for the geometry shown in Fig.~4(a) of the main text---where $J_{26}=J_c$ is the only nonzero interqubit coupling, and all intraqubit couplings are set to $J_0 = 20$~MHz---and obtain up to third order in $J_c$:

\begin{equation}
 H_\text{eff} = \frac{J_c^2}{32J_0 }\begin{pmatrix}
 -3 - \frac{3}{4}\frac{J_c}{J_0} & 0 & 0 & 0 \\
        0 & -11 - \frac{39}{4}\frac{J_c}{J_0} & 0 & 0 \\
        0 & 0 & -11 - \frac{39}{4}\frac{J_c}{J_0} & 0 \\
        0 & 0 & 0 & \frac{7}{3} + \frac{5}{4}\frac{J_c}{J_0}
    \end{pmatrix},
    \label{eq:supp_matrix}
\end{equation}
where our basis is  $|00\rangle, |01\rangle, |10\rangle, |11\rangle$. This is equivalent to Eq.~(4) in the main text when projected onto the basis of the computational Pauli matrices. We then compare this analytical result with the exact numerical simulation of the two-qubit gate in Fig.~\ref{fig:SchrifferWolff}, where we compute the Makhlin invariants~\cite{makhlin_nonlocal_2002} of the gate at each time step to find a gate equivalent to CNOT up to local unitaries. We find that a second-order Schrieffer-Wolff transformation captures the general evolution to CNOT and a third-order transformation correctly reproduces the CNOT gate time as predicted by numerics, although leakage oscillations which significantly impact the fidelity are visible in the numerical results. The need to go to third order in the Schrieffer-Wolff transformation is due to the fact our coupling $J_c$ must not be too small compared to the intraqubit coupling $J_0$ in order to achieve a reasonable gate time.

The Makhlin invariants of a two-qubit unitary $U$ are computed by converting $U$ into the Bell basis via the transformation $M_B = Q^\dagger U Q$, where
\begin{equation}
 Q = \frac{1}{\sqrt{2}}
    \begin{pmatrix}
        1 & 0 & 0 & i \\
        0 & i & 1 & 0 \\
        0 & i & -1 & 0 \\
        1 & 0 & 0 & -i 
    \end{pmatrix}.
\end{equation}
The invariants are given as $G_1 = \textrm{tr}(m)^2 \det U^\dagger / 16 $ and $G_2 = (\textrm{tr}(m)^2 - \textrm{tr}(m^2)) \det U^\dagger / 4 $, where $m = M^T_B M_B$. For the CNOT gate, the Makhlin invariants are $G_1 = 0$ and $G_2 = 1$.

\begin{figure}
    \centering
    \includegraphics[width=0.32\linewidth]{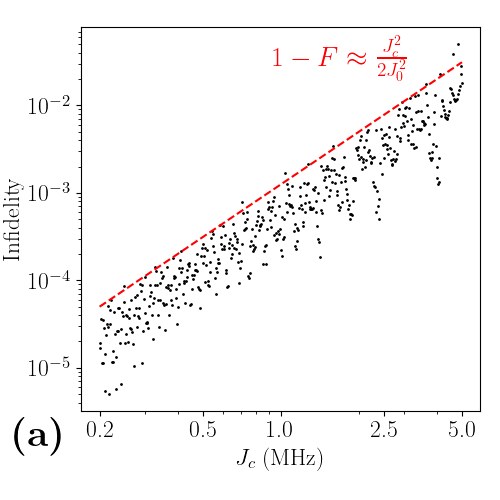}
    \includegraphics[width=0.32\linewidth]{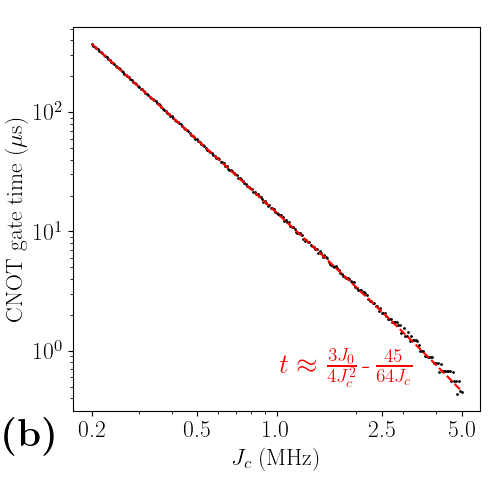}
    \includegraphics[width=0.32\linewidth]{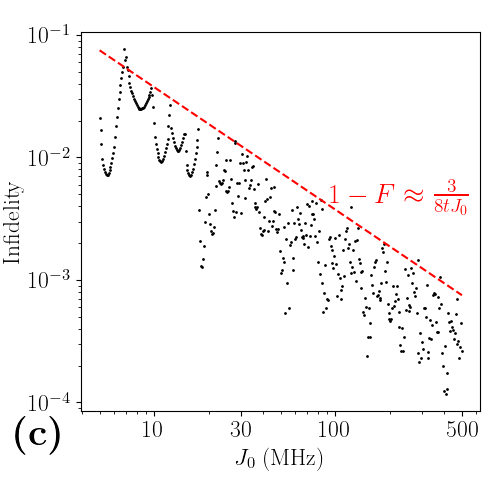}
    \caption{Characterization of the two-qubit gate time and intrinsic fidelity for various values of $J_0$ and $J_c$. (a) The intrinsic fidelity of the CNOT gate for a set value of $J_0 = 20$~MHz. The exact value of $F$ fluctuates across a wide band due to leakage. The red line represents a lower bound on the intrinsic fidelity.~ (b) The numerically found optimal gate times for the same value of $J_0 = 20$~MHz. (c) Holding the gate time constant at $t = 1$ $\mu$s, the fidelity over a range of values of $J_0$. $J_c = \sqrt{3J_0/4\cdot (1\mu\textrm{s})}$ in order to fix the expected gate time.}
    \label{fig:2q_fs_and_ts}
\end{figure}

We simulate CNOT gates over several values of $J_0$ and $J_c$ and find that a lower bound estimate of the intrinsic fidelity is given by the relationship $F \approx 1 - \frac{J_c^2}{2 J_0^2}$. The time of the gate can be estimated as $t_\textrm{CNOT} \approx \frac{3J_0} {4J_c^2} - \frac{45}{64J_c}$, plus the time of the significantly faster single-qubit unitaries. The numerical results illustrating these trends are shown in Fig.~\ref{fig:2q_fs_and_ts}.

\section{Effective Hamiltonians of interqubit couplings}

In all, there are sixteen possible interqubit exchange couplings. Because the matrix elements of the interqubit exchange interactions between computational states are zero, all effective two-qubit interactions are of order $J_c^2/J_0$. We perform a Schrieffer-Wolff transformation from turning on a single interqubit exchange coupling and calculate each effective Hamiltonian to be
\[
\begin{array}{c|c}
\textrm{Exchange coupling $J_c$} 
  & \textrm{Effective Hamiltonian (units of $J_c^2/16J_0$)} \\ \hline 
J_{15} 
  & \textrm{Identity} \\
J_{16} 
  & 2\,\sigma_2^z \\
J_{17} 
  & \sqrt{3}\,\sigma_2^x - \sigma_2^z \\
J_{18} 
  & -\sqrt{3}\,\sigma_2^x - \sigma_2^z \\
J_{25} 
  & 2\,\sigma_1^z \\
J_{26} 
  & \bigl(-2\,\sigma_1^z - 2\,\sigma_2^z + 8\,\sigma_1^z\sigma_2^z\bigr)/3 \\
J_{27} 
  & \bigl(\sigma_2^z - 2\,\sigma_1^z - \sqrt{3}\,\sigma_2^x 
 + 4\sqrt{3}\,\sigma_1^z\sigma_2^x 
 - 4\,\sigma_1^z\sigma_2^z\bigr)/3 \\
J_{28} 
  & \bigl(\sigma_2^z - 2\,\sigma_1^z + \sqrt{3}\,\sigma_2^x 
 - 4\sqrt{3}\,\sigma_1^z\sigma_2^x 
 - 4\,\sigma_1^z\sigma_2^z\bigr)/3 \\
J_{35} 
  & \sqrt{3}\,\sigma_1^x - \sigma_1^z \\
J_{36} 
  & \bigl(\sigma_1^z - 2\,\sigma_2^z - \sqrt{3}\,\sigma_1^x 
 + 4\sqrt{3}\,\sigma_1^x\sigma_2^z 
 - 4\,\sigma_1^z\sigma_2^z\bigr)/3 \\
J_{37} 
  & \bigl(\sigma_2^z + \sigma_1^z 
 - \sqrt{3}\,\sigma_2^x - \sqrt{3}\,\sigma_1^x 
 - 2\sqrt{3}\,\sigma_1^x\sigma_2^z 
 - 2\sqrt{3}\,\sigma_1^z\sigma_2^x 
 + 6\,\sigma_1^x\sigma_2^x 
 + 2\,\sigma_1^z\sigma_2^z\bigr)/3 \\
J_{38} 
  & \bigl(\sigma_2^z + \sigma_1^z 
 + \sqrt{3}\,\sigma_2^x - \sqrt{3}\,\sigma_1^x 
 - 2\sqrt{3}\,\sigma_1^x\sigma_2^z 
 + 2\sqrt{3}\,\sigma_1^z\sigma_2^x 
 - 6\,\sigma_1^x\sigma_2^x 
 + 2\,\sigma_1^z\sigma_2^z\bigr)/3 \\
J_{45} 
  & -\sqrt{3}\,\sigma_1^x - \sigma_1^z \\
J_{46} 
  & \bigl(\sigma_1^z - 2\,\sigma_2^z + \sqrt{3}\,\sigma_1^x 
 - 4\sqrt{3}\,\sigma_1^x\sigma_2^z 
 - 4\,\sigma_1^z\sigma_2^z\bigr)/3 \\
J_{47} 
  & \bigl(\sigma_2^z + \sigma_1^z 
 - \sqrt{3}\,\sigma_2^x + \sqrt{3}\,\sigma_1^x 
 + 2\sqrt{3}\,\sigma_1^x\sigma_2^z 
 - 2\sqrt{3}\,\sigma_1^z\sigma_2^x 
 - 6\,\sigma_1^x\sigma_2^x 
 + 2\,\sigma_1^z\sigma_2^z\bigr)/3 \\
J_{48} 
  & \bigl(\sigma_2^z + \sigma_1^z 
 + \sqrt{3}\,\sigma_2^x + \sqrt{3}\,\sigma_1^x 
 + 2\sqrt{3}\,\sigma_1^x\sigma_2^z 
 + 2\sqrt{3}\,\sigma_1^z\sigma_2^x 
 + 6\,\sigma_1^x\sigma_2^x 
 + 2\,\sigma_1^z\sigma_2^z\bigr)/3 
\end{array}
\]

The two-qubit basis used for these expressions is $\{\ket{00}, \ket{01}, \ket{10}, \ket{11}\}$ where $S_{12} = S_{34} = 0$ for state $\ket{0}$ and $S_{12} = S_{34} = 1$ for state $\ket{1}$.

\end{document}